\documentclass[a4paper,12pt]{article}
\usepackage{amsfonts,amsthm,amsmath,amssymb,graphicx,hyperref,color,youngtab}

\newcommand{\be}{\begin{equation}}
\newcommand{\bea}{\begin{eqnarray}}

\newcommand{\ee}{\end{equation}}
\newcommand{\eea}{\end{eqnarray}}

\begin{document}

\makeatletter
\@addtoreset{equation}{section}
\makeatother
\renewcommand{\theequation}{\thesection.\arabic{equation}}

\rightline{}
\vspace{1.8truecm}

\vspace{15pt}


{\LARGE{  
\centerline{\bf Rotating Restricted Schur Polynomials} 
}}  

\vskip.5cm 

\thispagestyle{empty}
    {\large \bf 
\centerline{Nicholas Bornman$^{a,b}$\footnote{{\tt bornman.nick@gmail.com}}, 
Robert de Mello Koch$^a$\footnote{{\tt robert@neo.phys.wits.ac.za}}, }
\centerline{ and Laila Tribelhorn$^a$\footnote{{\tt laila.tribelhorn@gmail.com}}}}

\vspace{.4cm}
\centerline{{\it ${}^a$National Institute for Theoretical Physics,}}
\centerline{{\it School of Physics and Mandelstam Institute for Theoretical Physics,}}
\centerline{{\it University of Witwatersrand, Wits, 2050, } }
\centerline{{\it South Africa } }

\vspace{.4cm}
\centerline{{\it ${}^b$Perimeter Institute for Theoretical Physics,}}
\centerline{{\it  Waterloo, Ontario, N2L 2Y5,}}
\centerline{{\it Canada} }

\vspace{1.4truecm}

\thispagestyle{empty}

\centerline{\bf ABSTRACT}

\vskip.4cm 

Large $N$ but non-planar limits of ${\cal N}=4$ super Yang-Mills theory can be described using restricted Schur polynomials.
Previous investigations demonstrate that the action of the one loop dilatation operator on restricted Schur operators, with 
classical dimension of order $N$ and belonging to the $su(2)$ sector, is largely determined by the $su(2)$ ${\cal R}$ symmetry 
algebra as well as structural features of perturbative field theory. 
Studies presented so far have used the form of ${\cal R}$ symmetry generators when acting on small perturbations 
of half-BPS operators.
In this article, as a first step towards going beyond small perturbations of the half-BPS operators,  we explain how the exact action 
of symmetry generators on restricted Schur polynomials can be determined.

\setcounter{page}{0}
\setcounter{tocdepth}{2}

\newpage

\tableofcontents

\setcounter{footnote}{0}

\linespread{1.1}
\parskip 4pt

{}~
{}~

\section{Orientation}

Studies of the anomalous dimensions of restricted Schur polynomials have demonstrated the emergence of giant
graviton brane states from ${\cal N}=4$ super Yang-Mills 
theory\cite{de Mello Koch:2007uu,de Mello Koch:2007uv,Bekker:2007ea,Koch:2011hb,deMelloKoch:2012ck,Koch:2015pga}.
The operators dual to giant graviton branes have a bare dimension that grows as $N$ as we take the large $N$
limit\cite{Balasubramanian:2001nh,Corley:2001zk,Berenstein:2004kk}.
Summing only the planar diagrams does not capture the large $N$ limit of these correlators\cite{Balasubramanian:2001nh}. 
The matching of field theory operators with excited giant graviton branes reproduces a rather non-trivial constraint implied by 
the Gauss Law of the gauge theory on the worldvolume of a giant graviton\cite{Balasubramanian:2004nb}.
An alternative approach, employing a collective coordinate description, has been developed 
in\cite{Berenstein:2013md,Berenstein:2013eya,Berenstein:2014pma,Berenstein:2014isa,Berenstein:2014zxa}.
The two approaches are in good agreement with each other and with studies in the dual string theory.

Studies of this type are important as they extend our understanding of gauge / gravity duality  into non-planar 
sectors of the field theory.
In these limits one is forced to sum higher genus ribbon graphs, so that these are inherently non-perturbative string
theory problems.
The results discussed above are limited to ${1\over 2}$-BPS and small deformations of ${1\over 2}$-BPS operators.
For these operators one can employ what is known as the ``distant corners approximation'', which supplies a dramatic
simplification of the action of the symmetric group.
One would like to extend these initial results to consider generic giant graviton branes and even operators with
dimensions that grow faster than $N$.
In this more general setting we can't justify the distant corners approximation and new methods are needed.

A promising approach to this problem is to explore how much can be obtained by exploiting the known symmetries
of the theory.
Indeed, using an $SU(2)$ ${\cal R}$-symmetry \cite{Koch:2013xaa} derived constraints that determine much of the structure 
of the higher loop contributions to the dilatation operator.
The basic idea is to compute the $su(2)$ generators in the displaced corners approximation and then to write down
the equations following from the requirement that these generators commute with the dilatation operator.
In this way a set of recursion relations for the dilatation operator are obtained.
By including additional knowledge - basically the requirement that the smallest possible anomalous dimension (for the
BPS operators) is zero - the recursion relations have an essentially unique solution.
If the symmetry generators are computed exactly, the equations following from the requirement that these generators commute 
with the dilatation operator would themselves be exact.
This set of equations would provide a starting point for studies that go beyond small deformations of the ${1\over 2}$-BPS
sector.  
Our main goal in this article is to compute the action of certain symmetry generators on restricted Schur polynomials, exactly.

In the next section of this article, we will derive exact expressions for the matrix elements of the generators we are
interested in.
Evaluating the formulas we obtain entails computing a trace over a product of projection operators.
This trace is easily evaluated in the displaced corners approximation and we quote the generators in this case.
The $su(2)$ generators in the displaced corners approximation were computed in \cite{Koch:2013xaa}.
The results for the $su(3)$ generators are new.
In section 3 we will tackle the computation of the exact $su(2)$ generators.
We are able to obtain identities that relate the traces over products of different projection operators.
With the help of these identities and a few well chosen initial traces, we are able to compute all such traces exactly.
The exact results are a rather simple modification of the displaced corners' generators.
In section 4 we consider the $su(3)$ generators, which involves three complex matrices.
A tensor product structure of the problem is made manifest.
This then allows the computation of the exact $su(3)$ generators, using the methods that worked for $su(2)$.
In section 5 we discuss our results.
In particular, our analysis makes it clear that the computation of any $su(n)$ generators rotating any of the fields
from which the restricted Schur operator is constructed, can be accomplished in exactly the same way.
We also sketch a number of interesting directions for further study. 

\section{Generators in the distant corners approximation}

${\cal N}=4$ super Yang-Mills theory includes 6 scalar fields $\phi_i$, that are $N\times N$ hermitian matrices transforming in 
the adjoint of the $U(N)$ gauge group.
The theory enjoys an $SO(6)$ global symmetry which rotates the scalars in the vector representation.
Form the complex combinations
\bea
  Z=\phi_1+i\phi_2,\qquad   Y=\phi_3+i\phi_4,\qquad   X=\phi_5+i\phi_6\, .
\eea
The global symmetry we study in this article is the $SU(3)\subset SO(6)$ group that rotates the above three complex fields.
In this section we will argue that the computation of the action of the $su(3)$ generators on restricted Schur polynomials
constructed using the above three complex fields amounts to the computation of the trace of a certain product of projection
operators. 
The relevant restricted Schur polynomials are given by
\bea
   \chi_{R,(r,s,t)\alpha\beta}(Z,Y,X)={1\over n!m!p!}\sum_{\sigma\in S_{n+m+p}}
    \chi_{R,(r,s,t)\alpha\beta}(\sigma){\rm Tr}(\sigma \, Z^{\otimes\, n}\otimes Y^{\otimes\, m}\otimes X^{\otimes\, p})\, .
\eea
The labels of the above operator are Young diagrams $R\vdash n+m+p$, $r\vdash n$, $s\vdash m$, $t\vdash p$
and multiplicity labels $\alpha,\beta$.
The definition is spelled out in complete detail in \cite{Bhattacharyya:2008rb}. 
We will ultimately focus on the case that $R$ has at most two rows or columns, which is dual to a system of two giant gravitons.
This case is the simplest to consider, since the multiplicity labels $\alpha$ and $\beta$ appearing above are not needed.
To illustrate the argument, it is enough to focus on the  $su(2)$ generators, as we do in the next section.
The results for the $su(3)$ sector, which we give below, can be obtained in exactly the same way.

\subsection{$SU(2)$ Rotations of Restricted Schur Polynomials}

We set $p=0$ and consider operators constructed using $n$ $Z$ fields and $m$ $Y$ fields.
The generators of the $su(2)$ algebra are as follows
\bea
   J_+={\rm Tr}\left(Y{d\over dZ}\right),\qquad    
   J_-={\rm Tr}\left(Z{d\over dY}\right),\qquad   
   J_3={\rm Tr}\left(Y{d\over dY}- Z{d\over dZ}\right)\, .
\eea
These generators close the usual algebra
\bea
   \big[J_+ ,J_-\big]=J_3,\qquad \big[J_3,J_\pm\big]=\pm 2 J_\pm\, .
\eea
We will make extensive use of the identity \cite{Bhattacharyya:2008xy}
\bea
 {\rm Tr}(\sigma Y^{\otimes\, m-1}\,\otimes\, Z^{\otimes\, n+1})=
 \sum_{T,(t^+,u^-)\vec{\nu}}{d_T (n+1)! (m-1)!\over d_{t^+} d_{u^-} (n+m)!}
  \chi_{T,(t^+,u^-)\vec{\nu}^*}(\sigma^{-1}) \chi_{T,(t^+,u^-)\vec{\nu}}(Z,Y)\cr
\label{complete}
\eea
where $t^+\vdash n+1,$ $u^-\vdash m-1$ and $T\vdash m+n$.

Consider first the action of $J_-$
\bea
 J_- \chi_{R,(r,s)\vec{\mu}}(Z,Y) &=& {\rm Tr}\left( Z{d\over dY}\right)\chi_{R,(r,s)\vec{\mu}}(Z,Y)\cr
   &=&{m\over n! m!}\sum_{\sigma\in S_{n+m}}{\rm Tr}_{(r,s)\vec{\mu}}\left(\Gamma^R(\sigma )\right)
      {\rm Tr}(\sigma Y^{\otimes\, m-1}\,\otimes\, Z^{\otimes\, n+1})\cr
   &=&\sum_{(t^+,u^-)\vec{\nu}}{n+1\over d_{t^+} d_{u^-} }
      {\rm Tr}_{R}(P_{R,(r,s)\vec{\mu}}P_{R,(t^+,u^-)\vec{\nu}^*})
      \chi_{R,(t^+,u^-)\vec{\nu}}(Z,Y)\, .\label{ident}
\eea
To move from the second last to the last line above, we have used the identity (\ref{complete}) and we
have performed the sum over $\sigma$ using the fundamental orthogonality relation
\bea
   \sum_{g\in {\cal G}}\Gamma^R(g)_{ab}\Gamma^S(g^{-1})_{cd}={|{\cal G}|\over d_R}\delta_{RS}\delta_{bc}\delta_{ad}\, ,
\eea
which holds for the matrix elements of the irreducible representations (irreps) of any group ${\cal G}$.
$R$ and $S$ label the irreps and $d_R$ is the dimension of irrep $R$.
Our notation in (\ref{ident}) is $\vec{\nu}=(\nu_1,\nu_2)$ and $\vec{\nu}^*=(\nu_2,\nu_1)$.
In bra-ket notation we have
\bea
  \langle T,(t^+,u^-),\vec{\nu} |{\rm Tr}\left( Z{d\over dY}\right)|R,(r,s),\vec{\mu}\rangle &=&\delta_{RT}
       {n+1\over d_{t^+} d_{u^-}} {\rm Tr}_{R}(P_{R,(r,s)\vec{\mu}}P_{R,(t^+,u^-)\vec{\nu}^*})\cr
&\equiv&\delta_{RT}
{\langle\chi_{R,(t^+,u^-)\vec{\nu}}^\dagger (Z,Y) {\rm Tr}\left( Z{d\over dY}\right)\chi_{R,(r,s)\vec{\mu}}(Z,Y) \rangle
\over
\langle \chi_{R,(t^+,u^-)\vec{\nu}}(Z,Y) \chi_{R,(t^+,u^-)\vec{\nu}}^\dagger (Z,Y) \rangle}
\, .\cr
&&
\eea
We want the action of the generators on normalized operators. Our operators obey
\bea
  \langle \chi_{R,(r,s)\vec{\mu}}(Z,Y)\chi_{T,(t,u)\vec{\nu}}^\dagger (Z,Y)\rangle
  =\delta_{RT}\delta_{rt}\delta_{su}\delta_{\vec{\mu}\vec{\nu}}{f_R{\rm hooks}_R\over {\rm hooks}_r {\rm hooks}_s}\, .
\eea
By rescaling 
$$
\chi_{R,(r,s)\vec{\mu}}(Z,Y) = \sqrt{f_R{\rm hooks}_R\over {\rm hooks}_r {\rm hooks}_s}O_{R,(r,s)\vec{\mu}}(Z,Y)
$$
we can get operators $O_{R,(r,s)\vec{\mu}}(Z,Y)$ with two point function equal to 1
$$
\langle O_{R,(r,s)\vec{\mu}}(Z,Y)O_{T,(t,u)\vec{\nu}}^\dagger (Z,Y)\rangle
  =\delta_{RT}\delta_{rt}\delta_{su}\delta_{\vec{\mu}\vec{\nu}}\, .
$$
Acting on normalized operators we have
\bea
   J_- O_{R,(r,s)\vec{\mu}}(Z,Y)=\sum_{T,(t^+,u^-)\vec{\nu}}
  (J_-)_{T,(t^+,u^-)\vec{\nu} \, ,\, R,(r,s)\vec{\mu}}O_{T,(t^+,u^-)\vec{\nu}}(Z,Y)
\eea
where
\bea
  &&(J_-)_{T,(t^+,u^-)\vec{\nu} \, ,\, R,(r,s)\vec{\mu}}\equiv
\langle O_{T,(t^+,u^-)\vec{\nu}}^\dagger (Z,Y) {\rm Tr}\left( Z{d\over dY}\right) O_{R,(r,s)\vec{\mu}}(Z,Y) \rangle\cr
&&\qquad  =
   \sum_i \delta_{RT} \delta_{t_i^{+\prime}r} 
\sqrt{{\rm hooks}_{t^+}{\rm hooks}_s\over {\rm hooks}_{r}{\rm hooks}_{u^-}}
   {\rm Tr}_{R}(P_{R,(r,s)\vec{\mu}}P_{R,(t^+,u^-)\vec{\nu}^*})\, .\label{matrixelements}
\eea
The computation is now reduced to evaluating the trace 
${\rm Tr}_{R}(P_{R,(r,s)\vec{\mu}}P_{R,(t^+,u^-)\vec{\nu}^*})$.
In the next section we will explain how to evaluate this trace in the displaced corners approximation, which is all
that is needed to study small deformations of the ${1\over 2}$-BPS sector of the theory.
Later we will explain how to evaluate the trace exactly.

\subsection{Distant Corners Approximation}

When the $S_m\subset S_{m+n}$ subgroup acts, it swaps labels belonging to $Y$ boxes in the Young-Yamanouchi
pattern filling $R$.
If these boxes are always well separated the action of $S_m$ is very simple: it acts as the identity if the permutation
permutes boxes in the same row, and simply swaps the entries of the boxes if they belong to different rows.
To ensure the validity of the approximation we must have $m\ll n$.
In this approximation, the labels $s,\alpha,\beta$ of a restricted Schur polynomial $\chi_{R,(r,s)\alpha\beta}(Z,Y)$
with a label $R$ that has $p$ rows or columns, can be traded for the labels of a state in an $SU(p)$ representation.
See \cite{Carlson:2011hy,Koch:2011hb} for further details.

Since we are considering restricted Schur polynomials labeled by a Young diagram $R$ with at most 2 rows, there is
no multiplicity index and we can use the $SU(2)$ state labels $j,j_3$ with $-j\le j_3\le j$ as usual.
Denote the length of row $i$ in $s$ (respectively $R,r$) by $s_i$ (respectively $R_i,r_i$).
The translation between the two labels is (see \cite{Koch:2011hb}, especially Appendix E)
\bea
   s_1&=&{m\over 2}+j\qquad s_2={m\over 2}-j\, ,\cr
   R_1&=&r_1+{m\over 2}+j_3\qquad R_2=r_2+{m\over 2}-j_3\, .
\eea
The computation of the relevant traces now reduces to the computation of $SU(2)$ Clebsch-Gordan coefficients.
For detailed examples of the computations required see \cite{Carlson:2011hy,Koch:2011hb} and especially
\cite{Koch:2013xaa} which computes precisely the traces that are used here.
The result is
\bea
  &&J_-O^{(n,m)}(r_1,j,j^3)=A_-O^{(n+1,m-1)}(r_1+1,j+{1\over 2},j_3-{1\over 2})\cr
                           &&+B_-O^{(n+1,m-1)}(r_1+1,j-{1\over 2},j_3-{1\over 2})
                         +C_-O^{(n+1,m-1)}(r_1,j+{1\over 2},j_3+{1\over 2})\cr
                           &&+D_-O^{(n+1,m-1)}(r_1,j-{1\over 2},j_3+{1\over 2})
\eea
where
\bea
  &&A_-=\sqrt{r_1}\sqrt{{m-2j\over 2}{2j+2\over 2j+1}}
       {j-j_3+1\over 2j+2}\, ,
\eea
\bea
 &&B_-=\sqrt{r_1}\sqrt{{m+2j+2\over 2}{2j\over 2j+1}}{j+j_3\over 2j}\, ,
\eea
\bea
  && C_-=\sqrt{r_2}\sqrt{{m-2j\over 2}{2j+2\over 2j+1}}{j+j_3+1\over 2j+2}\, ,
\eea
\bea
  && D_-=\sqrt{r_2}\sqrt{{m+2j+2\over 2}{2j\over 2j+1}}{j-j_3\over 2j}\, .
\eea
These generators do not close the correct $su(2)$ algebra, although it is correct to the leading order in ${m\over n}$,
as expected \cite{Koch:2013xaa}.

\subsection{$SU(3)$}

The non-trivial generators we want to study are
\bea
 {\rm Tr}\left( Z {d\over dY}\right)\quad
 {\rm Tr}\left( Z {d\over dX}\right)\quad
 {\rm Tr}\left( Y {d\over dZ}\right)\quad
 {\rm Tr}\left( Y {d\over dX}\right)\quad
 {\rm Tr}\left( X {d\over dY}\right)\quad
 {\rm Tr}\left( X {d\over dZ}\right).
\eea
The computation is almost exactly the same as for the $su(2)$ generators, so we will only point out what the differences
are and quote the final results.
First, the identity (\ref{complete}) must be upgraded.
For the case of three complex matrices the statement of the completeness of the restricted Schur polynomials is \cite{Koch:2012sf}
\bea
   {\rm Tr} (\sigma Z^{\otimes n}\otimes  Y^{\otimes m}\otimes X^{\otimes p})=
\sum_{R,(r,s,t)\vec{\alpha}\vec{\beta}}
{d_R n! m! p!\over d_{r}d_{s}d_{t} (n+m+p)!}
\chi_{R,(r,s,t)\vec{\alpha}\vec{\beta}}(\sigma^{-1})
\chi_{R,(r,s,t)\vec{\beta}\vec{\alpha}}(Z,Y,X)\, .\cr
\label{thefullthing}
\eea
Consider the generator
\bea
&& {\rm Tr}\left( X {d\over dY}\right)
   \chi_{R,(r,s,t)\vec{\alpha}\vec{\beta}}(Z,Y,X)
= {m \over n!m!p!}\sum_{\sigma\in S_{n+m+p}}
{\rm Tr}(P_{R,(r,s,t)\vec{\alpha}\vec{\beta}}\Gamma^R(\sigma))\times\cr
&&\qquad\times\sum_{T,(t_1,t_2,t_3)\vec{\delta}\vec{\gamma}}
{d_T n! (m-1)! (p+1)! \over d_{t_1}d_{t_2}d_{t_3}(n+m+p)!}
\chi_{T,(t_1,t_2,t_3)\vec{\delta}\vec{\gamma}}(\sigma^{-1})
\chi_{T,(t_1,t_2,t_3)\vec{\gamma}\vec{\delta}}(Z,Y,X)\, .\cr
&&\label{firstgenerator}
\eea
In the above, $T\vdash n+m+p$, $t_1\vdash n$, $t_2\vdash m-1$ and $t_3\vdash p+1$.
Carrying out the sum over $\sigma$ in (\ref{firstgenerator}) using the completeness relation, we find
\bea
&& {\rm Tr}\left( X {d\over dY}\right)
       \chi_{R,(r_1,r_2,r_3)\vec{\alpha}\vec{\beta}}(Z,Y,X)
     ={m \over n!m!p!}\sum_{(t_1,t_2,t_3)\vec{\delta}\vec{\gamma}}\cr
&&{(m-1)! n! (p+1)! \over d_{t_1}d_{t_2}d_{t_3}}
{\rm Tr}\left( P_{R,(r_1,r_2,r_3)\vec{\alpha}\vec{\beta}}P_{R,(t_1,t_2,t_3)\vec{\delta}\vec{\gamma}}\right)
\chi_{R,(t_1,t_2,t_3)\vec{\gamma}\vec{\delta}}(Z,Y,X)\, .
\eea
To evaluate this generator, we need to evaluate the trace
\bea
{\rm Tr}\left( P_{R,(\vec{r})\vec{\alpha}\vec{\beta}}P_{R,(\vec{t})\vec{\delta}\vec{\gamma}}\right)\, .
\eea
It is again simplest to focus on the two row example and to use the distant corners approximation.
The translation of the restricted Schur polynomial $\chi_{R,(r,s,t)}(Z,Y,X)$ to $SU(2)$ state labels is as follows
\bea
   t_1&=&{p\over 2}+k\qquad t_2={p\over 2}-k\, ,\cr
   s_1&=&{m\over 2}+j\qquad s_2={m\over 2}-j\, ,\cr
   R_1&=&r_1+{m+p\over 2}+j_3+k_3\qquad R_2=r_2+{m+p\over 2}-j_3-k_3\, .\label{translate}
\eea
There is a point worth noting here: the above labels may appear to be over complete. 
Indeed, given $n,m,p$ as well as $r,k,j,k_3+j_3$ we can reconstruct the Young diagram labels $R,r,s$ and $t$.
It seems that we need only the sum $k_3+j_3$ and not the individual values $j_3,k_3$.
The point is that, even when $R$ has two rows, when we restrict $S_{p+m+n}$ to $S_p\times S_m\times S_n$
we do need a multiplicity label.
Specifying $k_3$ and $j_3$ independently resolves the multiplicity.
The simplest way to see this is to note that we can first restrict $S_{p+m+n}$ to $S_p\times S_{m+n}$ without
multiplicity, and then restrict $S_{m+n}$ to $S_m\times S_n$, again without multiplicity.
The first restriction introduces $(k,k_3)$ and the second $(j,j_3)$.

In the distant corners approximation, the computation of the traces needed to compute the generators is again reduced to 
the computation of $SU(2)$ Clebsch-Gordan coefficients.
Since the boxes associated to the $X$ fields (organized by Young diagram $t\vdash p$) are removed first,
the generators ${\rm Tr}\left( Z{d\over dY}\right)$ and ${\rm Tr}\left( Y{d\over dZ}\right)$ are unchanged
from the formulas we obtained above.
If we compute the action of ${\rm Tr}\left( X{d\over dY}\right)$, the action of any other generator can be
computed by taking the hermitian conjugate or by using the $su(3)$ algebra.
Consequently, we only need (and quote) the action of ${\rm Tr}\left( X{d\over dY}\right)$.
We find
\bea
{\rm Tr}\left( X{d\over dY}\right)O^{(n,m,p)}_{R,r,j,j_3,k,k_3}\cr
={j+j_3\over 2j}{k+k_3+1\over 2k+1}
\sqrt{\Big({m\over 2}+j+1\Big){2j\over 2j+1}}
\sqrt{\Big({p\over 2}+k+2\Big){2k+1\over 2k+2}}
O^{(n,m-1,p+1)}_{R,r,j-{1\over 2},j_3-{1\over 2},k+{1\over 2},k_3+{1\over 2}}\cr
+{j+j_3\over 2j}{k-k_3\over 2k+1}
\sqrt{\Big({m\over 2}+j+1\Big){2j\over 2j+1}}
\sqrt{\Big({p\over 2}-k+1\Big){2k+1\over 2k}}
O^{(n,m-1,p+1)}_{R,r,j-{1\over 2},j_3-{1\over 2},k-{1\over 2},k_3+{1\over 2}}\cr
+{j-j_3+1\over 2j+2}{k+k_3+1\over 2k+1}
\sqrt{\Big({m\over 2}-j\Big){2j+2\over 2j+1}}
\sqrt{\Big({p\over 2}+k+2\Big){2k+1\over 2k+2}}
O^{(n,m-1,p+1)}_{R,r,j+{1\over 2},j_3-{1\over 2},k+{1\over 2},k_3+{1\over 2}}\cr
+{j-j_3+1\over 2j+2}{k-k_3\over 2k+1}
\sqrt{\Big({m\over 2}-j\Big){2j+2\over 2j+1}}
\sqrt{\Big({p\over 2}-k+1\Big){2k+1\over 2k}}
O^{(n,m-1,p+1)}_{R,r,j+{1\over 2},j_3-{1\over 2},k-{1\over 2},k_3+{1\over 2}}\cr
+{j-j_3\over 2j}{k-k_3+1\over 2k+1}
\sqrt{\Big({m\over 2}+j+1\Big){2j\over 2j+1}}
\sqrt{\Big({p\over 2}+k+2\Big){2k+1\over 2k+2}}
O^{(n,m-1,p+1)}_{R,r,j-{1\over 2},j_3+{1\over 2},k+{1\over 2},k_3-{1\over 2}}\cr
+{j-j_3\over 2j}{k+k_3\over 2k+1}
\sqrt{\Big({m\over 2}+j+1\Big){2j\over 2j+1}}
\sqrt{\Big({p\over 2}-k+1\Big){2k+1\over 2k}}
O^{(n,m-1,p+1)}_{R,r,j-{1\over 2},j_3+{1\over 2},k-{1\over 2},k_3-{1\over 2}}\cr
+{j+j_3+1\over 2j+2}{k-k_3+1\over 2k+1}
\sqrt{\Big({m\over 2}-j\Big){2j+2\over 2j+1}}
\sqrt{\Big({p\over 2}+k+2\Big){2k+1\over 2k+2}}
O^{(n,m-1,p+1)}_{R,r,j+{1\over 2},j_3+{1\over 2},k+{1\over 2},k_3-{1\over 2}}\cr
+{j+j_3+1\over 2j+2}{k+k_3\over 2k+1}
\sqrt{\Big({m\over 2}-j\Big){2j+2\over 2j+1}}
\sqrt{\Big({p\over 2}-k+1\Big){2k+1\over 2k}}
O^{(n,m-1,p+1)}_{R,r,j+{1\over 2},j_3+{1\over 2},k-{1\over 2},k_3-{1\over 2}}\, .\cr
\eea
Once again, the $su(3)$ algebra is not obeyed exactly. 
It is obeyed to leading order in ${m\over n}$ and ${p\over n}$, which is exactly what we'd expect.

\section{Analytic computation of exact $su(2)$ generators}

In this section we will compute the matrix elements of $J_-$ exactly.
The basic insight we exploit is that, in the case of two rows where we don't need the multiplicity labels, the
projector can be written as
\bea
   P_{R,(r,s)}={1\over m!}\sum_{\sigma\in S_m}\chi_s (\sigma )\Gamma_R(\sigma)\, .
\eea
The above projector must be restricted to act on the space of states obtained by labeling the $m$ boxes that must be peeled off 
of $R$ to leave $r$.
In the next four sections we will derive a collection of identities that, when taken together, allow us to compute the
trace appearing in the $su(2)$ generators exactly.
We are again focusing on operators labeled by Young diagrams with at most two rows or columns.

\subsection{Restricted Character Formula} \label{restchar}

Our computation involves a restricted character that is easily evaluated using the methods developed in 
\cite{Bhattacharyya:2008xy}.
See also \cite{Mattioli:2016eyp} for a closely related recent discussion.
Introduce indices with the following ranges
\bea
   I,J=1,2,...,m+n,\qquad \alpha,\beta = m+1,m+2,...,m+n,\qquad a,b=1,2,...,m.\nonumber
\eea
We will compute the restricted character
\bea
   \chi_{R,(r,s)}(\,(a,\alpha)\,) = {\rm Tr}(P_{R,(r,s)}(a,\alpha )\,)\, .
\eea
Denote the row lengths of these Young diagrams by $R_1,R_2$, $r_1,r_2$ and $s_1,s_2$.
By $\sum_{IJ} (I,J)$ we mean the sum of all distinct two cycles.
For $S_n$ this sum runs over $n(n-1)/2$ terms.
It is easy to establish the identity
\bea
   \sum_{a,\alpha} (a,\alpha) = \sum_{IJ}(I,J)-\sum_{ab}(a,b)-\sum_{\alpha\beta}(\alpha,\beta)\, .
   \label{basicident}
\eea
Recall that the sum over all two cycles is a Casimir with eigenvalue equal to the number of row pairs minus the number of column 
pairs \cite{chen}.
As a result, since our Young diagrams have two rows, it follows that
\bea
   \sum_{IJ}{\rm Tr}(P_{R,(r,s)} (I,J)\,)={R_1 (R_1-1)\over 2}+{R_2 (R_2-1)\over 2}-R_2\, ,
\eea
\bea
   \sum_{\alpha\beta}{\rm Tr}(P_{R,(r,s)} (\alpha,\beta)\,)={r_1 (r_1-1)\over 2}+{r_2 (r_2-1)\over 2}-r_2\, ,
\eea
\bea
   \sum_{ab}{\rm Tr}(P_{R,(r,s)} (a,b)\,)={s_1 (s_1-1)\over 2}+{s_2 (s_2-1)\over 2}-s_2\, .
\eea
Taking the trace of (\ref{basicident}) it follows that
\bea
   \sum_{a,\alpha}\chi_{R,(r,s)}(\,(a,\alpha)\,)&=& nm \chi_{R,(r,s)}(\,(m,m+1)\,)\cr
&=&{2j_3^2 -2j(j+1)+2j_3 (r_1-r_2+1)+m(r_1+r_2)\over 2} d_r d_s \cr
&&
\eea
which is the formula we were after.

\subsection{A Simple Trace}
 
The identities we derive in the next two subsections after this one give the answer for the trace over a sum of projectors.
In this section we will compute a trace that is simple enough to evaluate exactly.
This trace, together with the identities we obtain, allow us to determine all of the remaining traces.

We want to compute
\bea
   T= {\rm Tr}(P_{R,(r,s)}P_{R,(r^+,s^-)})= \sum_i \langle R,(r,s); i|P_{R,(r^+,s^-)}|R,(r,s);i\rangle
\eea
with $R\vdash n+m$, $r\vdash n$, $r^+\vdash n+1$, $s\vdash m$ and $s^-\vdash m-1$.
One way of embedding $S_n$ irrep $r$ within $S_{n+m}$ irrep $R$ is to remove boxes from $R$ to obtain $r$.
Assume that we remove $m_1$ boxes from the first row of $R$ and $m_2$ boxes from the second row of $R$.
The first basic trace we will compute assumes that $r^+$ is given by adding one box to the second row of $r$.
The Young diagram labeling $s$ is a single row of $m$ boxes.
The Young diagram labeling $s^-$ is a single row of $m-1$ boxes.
It is clear that the only states that participate on the right hand side of the above equation have box $m$
(the last box of $s$ which is the first box of $r^+$) in the second row of $r^+$.
Further, only the subspace of $s$ corresponding to $s^-$ contributes.
Thus, we need to do the sum over $i$ making sure that we arrange two things: box $m$ must sit in the second
row and we must project to $s^-$.
Projecting to $s^-$ is easy: we know how to construct a projector that will accomplish this.
To fix box $m$, note that the last box of $s$ can be in the first row or the second row.
By fixing the content of this box we can ensure its in the second row.
This content is measured by the Jucys-Murphy element which lives in the $S_m$ group algebra \cite{chen}.
Using this element we can construct the operator
\bea
\left[ {r_1-\sum_{i=m+1}^{m+n} (m,i)\over r_1-r_2+1}\right]
\eea
which gives $1$ when acting on states for which the last box of $r^+$ is in the second row and zero
when the last box of $r^+$ is in the first row.
Clearly then
\bea
   T&=& \sum_i \langle R,(r,s); i|
         \left[ {r_1-\sum_{i=m+1}^{m+n} (m,i)\over r_1-r_2+1}\right]\left[ {1\over (m-1)!}\sum_{\sigma\in S_{m-1}}
          \chi_{s^-}(\sigma)\Gamma_R(\sigma)\right] |R,(r,s);i\rangle\cr
&=& \sum_i \langle R,(r,s); i|
         \left[ {r_1-\sum_{i=m+1}^{m+n} (m,i)\over r_1-r_2+1}\right]\left[ {1\over (m-1)!}\sum_{\sigma\in S_{m-1}}
          \Gamma_R(\sigma)\right] |R,(r,s);i\rangle\cr
&=& \sum_i \langle R,(r,s); i|
         \left[ {r_1-\sum_{i=m+1}^{m+n} (m,i)\over r_1-r_2+1}\right]\left[ {1\over (m-1)!}\sum_{\sigma\in S_{m-1}}
          \Gamma_s(\sigma)\right] |R,(r,s);i\rangle\cr
&=& \sum_i \langle R,(r,s); i|
         \left[ {r_1-\sum_{i=m+1}^{m+n} (m,i)\over r_1-r_2+1}\right] |R,(r,s);i\rangle\cr
&=& {r_1 d_r d_s\over r_1-r_2+1}-n{\chi_{R,(r,s)}(\,(m,m+1)\,)\over r_1-r_2+1}\cr
&=& {m-2j_3\over 4m}\left[ 2 + {m+2j_3\over r_1-r_2+1}\right] d_r d_s
\eea
where to obtain the second line we use $\chi_{s^-}(\sigma)=1$, to obtain the third line we use the fact that
state $|R,(r,s);i\rangle$ belongs to an irrep of $S_n\times S_m$, to obtain the fourth line we use the fact
that in irrep $s$ we have $\Gamma_s(\sigma)=1$ and to obtain the last line use the formula for the restricted character derived 
in the previous section.

The second basic trace that we will need arises when $r^+$ is given by adding one box to row 1 of $r$.
The second basic trace is easily computed, in exactly the same way, to be
\bea
   T=  \left[ {m+2j_3\over 2m}- {(m^2-4(j_3)^2)\over 4m(r_1-r_2+1)}\right] d_r d_s\, .
\eea

\subsection{Identities for the trace of a sum of projectors}

We will derive two identities in this section.
To obtain the first identity, assume $r^+$ is given by adding a box to row 2 of $r$.
It is trivial to obtain the formula for $r^+$ given by adding a box to row 1.
The generic $s$ subduces two possible $s^-$: $s_1^-$ and $s_2^-$, where
$s_1^-$ is obtained by dropping a box from row 1 of $s$ and $s_2^-$ is obtained by dropping a box from row 2 of $s$.
It is now rather straightforward to compute (we use $s= s^-_1\oplus s^-_2$ below)
\bea
   T&=&\sum_{K=1}^2 {\rm Tr}\left(P_{R,(r,s)}P_{R,(r^+,s^-_K)}\right)\cr
&=& \sum_i \langle R,(r,s); i|
         \left[ {r_1-\sum_{i=m+1}^{m+n} (m,i)\over r_1-r_2+1}\right]\times\cr
        &&\quad\times \sum_K \left[ {1\over (m-1)!}\sum_{\sigma\in S_{m-1}}
          \chi_{s^-_K}(\sigma)\Gamma_R(\sigma)\right] |R,(r,s);i\rangle\cr
&=& \sum_i \langle R,(r,s); i|
         \left[ {r_1-\sum_{i=m+1}^{m+n} (m,i)\over r_1-r_2+1}\right]\times\cr
        &&\quad\times \left[ {1\over m!}\sum_{\sigma\in S_{m}}
          \chi_{s}(\sigma)\Gamma_R(\sigma)\right] |R,(r,s);i\rangle\cr
&=& \sum_i \langle R,(r,s); i|
         \left[ {r_1-\sum_{i=m+1}^{m+n} (m,i)\over r_1-r_2+1}\right] |R,(r,s);i\rangle\cr
&=&\left[ {m-2j_3\over 2m}+{2j(j+1)-2j_3^2-m\over 2m(r_1-r_2+1)}\right] d_r d_s\, .
\eea
This is the first sum identity we wanted to prove.

For the second sum identity, we start by noting that the generic $s^-\vdash m-1$ can be subduced, by two possible 
representations $s_1$ and $s_2$.
If you drop a box from row 1 of $s_1$ you get $s^-$ and if you drop a box from row 2 of $s_2$ you get $s^-$.
We want to compute
\bea
T=\sum_{J=1}^2 {\rm Tr} (P_{R,(r,s_J)}P_{R,(r^+,s^-)})\, .
\eea
We can easily extend the sum above to a sum over all the possible $s$: since the only irreps that can subduce $s^-$
are $s_1$ and $s_2$, the additional terms all vanish. 
Now,
\bea
   \sum_s P_{R,(r,s)}
\eea
projects us from $R$ to $r$.
Thus, the product $P_{R,(r,s_J)}P_{R,(r^+,s^-)}$ projects us to $r\oplus {\tiny \yng(1)}\oplus s^-$ and hence
\bea
   T=\sum_{J=1}^2 {\rm Tr} (P_{R,(r,s_J)}P_{R,(r^+,s^-)})=d_r d_{s^-}\, .
\eea

\subsection{Example of how the trace identities are used}

Using the two basic traces and the two basic sum identities, we can compute any trace we want.
This is easily illustrated with an example.
Imagine we remove 3 boxes from row 1 and 2 boxes from row 2 of $R$ to get $r$.
The possible projectors are
\bea
   P_1=P_{\tiny \yng(8,4),\yng(5,2),\yng(5)}\, ,
\eea
\bea
   P_2=P_{\tiny \yng(8,4),\yng(5,2),\yng(4,1)}\, ,
\eea
\bea
   P_3=P_{\tiny \yng(8,4),\yng(5,2),\yng(3,2)}\, .
\eea
For the second set of projectors imagine that we remove 3 boxes from row 1 and 1 box from row 2 of $R$ to get $r^+$.
The possible projectors are
\bea
   P_A=P_{\tiny \yng(8,4),\yng(5,3),\yng(4)}\, ,
\eea
\bea
   P_B=P_{\tiny \yng(8,4),\yng(5,3),\yng(3,1)}\, .
\eea
We compute ${\rm Tr}(P_1 P_A)$ using our basic trace.
Given this, we compute ${\rm Tr}(P_2 P_A)$ using our second sum identity.
Given this, we compute ${\rm Tr}(P_2 P_B)$ using our first sum identity.
Given this, we compute ${\rm Tr}(P_3 P_B)$ using our second sum identity.
Since the trace of any other product of projectors vanishes, this is the complete set.

\subsection{Exact results for the $su(2)$ generators}
 
We will shift from the $R,(r,s)$ notation to the $r_1,j,j_3,n,m$ notation.
The exact results for the traces we need are
\bea
   {\rm Tr}\left(P(r_1,j,j_3,n,m)P(r_1,j-{1\over 2},j_3+{1\over 2},n+1,m-1)\right)\cr
=\left[{j-j_3\over 2j}+{j^2-j_3^2\over 2j (r_1-r_2+1)}\right]d_rd_{s^-}\, ,
\label{fsu2}
\eea
\bea
   {\rm Tr}\left(P(r_1,j,j_3,n,m)P(r_1,j+{1\over 2},j_3+{1\over 2},n+1,m-1)\right)\cr
=\left[{j+j_3+1\over 2j+2}-{(j+1)^2-j_3^2\over (2j+2) (r_1-r_2+1)}\right]d_rd_{s^-}
\, ,\label{ssu2}
\eea
\bea
   {\rm Tr}\left(P(r_1,j,j_3,n,m)P(r_1+1,j-{1\over 2},j_3-{1\over 2},n+1,m-1)\right)\cr
=\left[{j+j_3\over 2j}-{j^2-j_3^2\over 2j (r_1-r_2+1)}\right]d_rd_{s^-}\, ,\label{tsu2}
\eea
\bea
   {\rm Tr}\left(P(r_1,j,j_3,n,m)P(r_1+1,j+{1\over 2},j_3-{1\over 2},n+1,m-1)\right)\cr
=\left[{j-j_3+1\over 2j+2}+{(j+1)^2-j_3^2\over (2j+2) (r_1-r_2+1)}\right]d_rd_{s^-}\, .\label{lsu2}
\eea
These traces are all that is needed for the construction of the exact matrix elements of $J_-$.
These matrix elements are
\bea
  &&J_-O^{(n,m)}(r_1,j,j_3)=A_-O^{(n+1,m-1)}(r_1+1,j+{1\over 2},j_3-{1\over 2})\cr
                         &&+B_-O^{(n+1,m-1)}(r_1+1,j-{1\over 2},j_3-{1\over 2})
                         +C_-O^{(n+1,m-1)}(r_1,j+{1\over 2},j_3+{1\over 2})\cr
                          &&+D_-O^{(n+1,m-1)}(r_1,j-{1\over 2},j_3+{1\over 2})
\eea
where
\bea
  &&A_-=\sqrt{(r_1+2)(r_1-r_2+1)\over (r_1-r_2+2)}\sqrt{{m-2j\over 2}{2j+2\over 2j+1}}
       \left[{j-j_3+1\over 2j+2}+{(j+1)^2-j_3^2\over (2j+2) (r_1-r_2+1)}\right]\, ,\cr
&&
\eea
\bea
 &&B_-=\sqrt{(r_1+2)(r_1-r_2+1)\over (r_1-r_2+2)}\sqrt{{m+2j+2\over 2}{2j\over 2j+1}}
\left[{j+j_3\over 2j}-{j^2-j_3^2\over 2j (r_1-r_2+1)}\right]\, ,\cr
&&
\eea
\bea
  && C_-=\sqrt{(r_2+1)(r_1-r_2+1)\over (r_1-r_2)}\sqrt{{m-2j\over 2}{2j+2\over 2j+1}}
  \left[{j+j_3+1\over 2j+2}-{(j+1)^2-j_3^2\over (2j+2) (r_1-r_2+1)}\right]\, ,\cr
  &&
\eea
\bea
  && D_-=\sqrt{(r_2+1)(r_1-r_2+1)\over (r_1-r_2)}\sqrt{{m+2j+2\over 2}{2j\over 2j+1}}
                  \left[{j-j_3\over 2j}+{j^2-j_3^2\over 2j (r_1-r_2+1)}\right]\, .\cr
&&
\eea
The action of $J_3$ is particularly simple
\bea
J_3\, O^{(n,m)}(r_1,j,j_3)=(m-n)\, O^{(n,m)}(r_1,j,j_3)\, .
\eea
To get the matrix elements of $J_+$ we can dagger the above expressions.
In a bra-ket notation the relevant formula is
\bea
\langle n+1,r^+,m-1,j',j_3'|J_+|n,r,m,j,j_3\rangle
=\cr
\langle n,r,m,j,j_3|J_-|n+1,r^+,m-1,j',j_3',p+1\rangle\, .
\eea
The above generators do close the correct $su(2)$ algebra, which is a good check.
Further, in the distant corners approximation we have $r_1-r_2\gg 1$.
In this limit we see that the above generators reduce to the formulas we obtained in the displaced corners limit.
Finally, for $n=3$ and $m=2$ we have checked by hand, using explicit examples, that the above action is indeed correct.
This completes the derivation of the exact $su(2)$ generators.

\section{Analytic Computation of $su(3)$ generators}

In this section we compute the matrix elements of the $su(3)$ generators exactly.
For the $su(2)$ algebra, we only needed to compute $J_-$. 
We obtain $J_+$ by taking the hermitian conjugate of $J_-$ and the action of $J_3$ follows from the algebra.
For the $su(3)$ algebra we again only need to compute a single generator.
First, it follows that since the boxes associated to the $X$ fields (organized by Young diagram $t\vdash p$) are removed first,
the generators ${\rm Tr}\left( Z{d\over dY}\right)$ and ${\rm Tr}\left( Y{d\over dZ}\right)$ are unchanged
from the formulas we obtained for $su(2)$.
We will compute the action of ${\rm Tr}\left( X{d\over dY}\right)$.
The action of  ${\rm Tr}\left( Y{d\over dX}\right)$ follows from this by taking the hermitian conjugate and the action of any 
other generator can then be computed using the $su(3)$ algebra.
Consequently, we only need the action of ${\rm Tr}\left( X{d\over dY}\right)$.

The approach that worked for the $su(2)$ algebra can be used to compute the $su(3)$ generators.
In particular, there are again identities we can derive for traces over sums of projectors, and specific matrix
elements that we are able to compute exactly.
These identities then completely determine the traces and hence the generator in general.
In the end however, the  final results have been verified by checking that the generators reduce to the displaced corners
result, close the correct algebra and give the correct action for small values of $m$, $n$ and $p$ where the action can be
computed explicitly.

In what follows we freely move between the\footnote{Including $R/p$ as a label is equivalent to specifying $j_3$ and $k_3$ 
separately. See the comment after (\ref{translate}).} $R,R/p,(r,s,t)$ and $R,r,j,j_3,k,k_3$ labeling of our operators.
For concrete final results, to improve the clarity of our discussion, we will denote the number of $Z$s ($n$),
the number of $Y$s ($m$) and the number of $X$s ($p$) using a superscript.

\subsection{Tensor Product Structure of the Problem}

The states of irrep $(r,s,t)$ of $S_n\times S_m\times S_p$ are obtained in the carrier space of irrep $R$ of $S_{n+m+p}$,
by associating boxes inside $R$ with $r$ ($Z$ boxes), $s$ ($Y$ boxes) or $t$ ($X$ boxes).
These three species of boxes are filled independently as follows: $X$ boxes are filled with the labels $1,2,...,p$, $Y$ boxes with
$p+1,p+2,...,p+m$ and $Z$ boxes with $p+m+1,p+m+2,...,p+m+n$.
Each species is filled independently of the others.
It is clear that each vector belonging to this basis is the tensor product of a $Z$ vector, a $Y$ vector and an $X$ vector. 
The $Z$ boxes are already organized according to $r$. 
Thus, we can write our projector as
\bea
   P_{R,(r,s,t)}&=&{1\over m!}\sum_{\sigma\in S_m}\chi_s(\sigma )\Gamma_R(\sigma )\,
                          {1\over p!}\sum_{\tau \in S_p}\chi_t(\tau )\Gamma_R (\tau)\cr
                       &\equiv& {\bf 1}^R_r\otimes P^R_s\otimes P^R_t
\eea
where we have restricted this operator to the subspace spanned by states filled as we just described and $S_m$ acts on
$p+1,p+2,...,p+m$ while $S_p$ acts on $1,2,...,p$.
The tensor product structure is also evident in this projector.
To compute the trace
\bea
   {\rm Tr}(P^{n,m,p}_{R,(r,s,t)}\, P^{n,m-1,p+1}_{R,(r,s^-,t^+)})=
   {\rm Tr}({\bf 1}^R_r\otimes P^R_s\otimes P^R_t\, {\bf 1}^R_r\otimes P^R_{s^-}\otimes P^R_{t^+})
\eea
we can decompose $P_s^R$ into a sum of 2 projectors onto $S_{m-1}$ irreps
\bea
   P_s^R= \alpha_1 P_{s'_1}^R + \alpha_2 P_{s'_2}^R\label{ydecomp}
\eea
and $P_{t^+}^R$ into a sum of 2 projectors onto $S_{p}$ irreps
\bea
   P_{t^+}^R= \beta_1 P_{t^{+\prime}_1}^R + \beta_2 P_{t^{+\prime}_2}^R\, .\label{xdecomp}
\eea
Our notation uses $T'_i$ to denote the Young diagram obtained by dropping a box from row $i$ of $T$.
Assuming, for example, that $s_1'=s^-$ and $t^{+\prime}_2=t$ we find
\bea
   {\rm Tr}(P^{n,m,p}_{R,(r,s,t)}\, P^{n,m-1,p+1}_{R,(r,s^-,t^+)})=
   d_r d_{s^-}d_t\alpha_1\beta_2\, .
\eea
Clearly then, our traces are determined once we have understood how to decompose the projectors.
The problems (\ref{ydecomp}) and (\ref{xdecomp}) are independent of each other.
The problem of determining the decomposition (\ref{xdecomp}),  can be understood by studying the 
$S_{n+m+p}\to S_{n+m}\times S_p$ problem.
This shows that the results of the decomposition (\ref{xdecomp}) can be read straight from the $su(2)$ trace results of the 
previous section (formulas (\ref{fsu2}), (\ref{ssu2}), (\ref{tsu2}) and (\ref{lsu2})), after replacing $r_1-r_2\to r_1-r_2+2j_3$.
Thus, we only need to study the decomposition (\ref{ydecomp}).

\subsection{Identities for the trace of a sum of projectors}

Since the two decompositions (\ref{xdecomp}) and (\ref{ydecomp}) are independent of each other, when studying the
problem (\ref{ydecomp}) we can simplify things enormously by setting $p=0$.
Consequently, we need only study the problem of computing
\bea
   {\rm Tr}(P^{n,m,0}_{R,(r,s)}\, P^{n,m-1,1}_{R,(r,s^-,{\tiny \yng(1)})})
\eea
This collection of traces can again be computed by deriving two trace identities as we did for the $su(2)$ problem.
The simplest identity, derived exactly as for the $su(2)$ case, is
\bea
   \sum_s {\rm Tr}(P^{n,m,0}_{R,(r,s)}\, P^{n,m-1,1}_{R,(r,s^-,{\tiny \yng(1)})})=d_r d_{s^-}\label{frstsu3}
\eea
The second identity we need follows from considering
\bea
   \sum_{s^-}{\rm Tr}(P^{n,m,0}_{R,(r,s)}\, P^{n,m-1,1}_{R,(r,s^-,{\tiny \yng(1)})})
   ={\rm Tr}(P^{n,m,0}_{R,(r,s)}\, \hat{C})
\eea
Assume for concreteness that $\yng(1)$ in $P^{n,m-1,1}_{R,(r,s^-,{\tiny \yng(1)})}$ sits in the first row of $R$.
On the LHS of the above equation, since we sum over all $s^-$, the sum $\sum_{s^-}P^{n,m-1,1}_{R,(r,s^-,{\tiny \yng(1)})}$
projects the $Z$ boxes to $r$ and the first box (corresponding to $\yng(1)$) to the right most box of the first row of $R$.
Since $P^{n,m,0}_{R,(r,s)}$ already projects the $Z$ boxes to $r$, the net affect is that 
$\sum_{s^-}P^{n,m-1,1}_{R,(r,s^-,{\tiny \yng(1)})}$ projects $\yng(1)$ to the right most box of the first row of $R$.
Arguing exactly as we did for the $su(2)$ problem, we can accomplish this using a Jucys-Murphy element.
Denote the number of boxes removed from row $i$ of $R$ to obtain $r$ by $m_i$.
A simple argument now shows that we can write $\hat{C}$ as follows
\bea
   \hat{C}={\sum_{i=2}^{m+n} (i,1) - (r_2+m_2-2)\over r_1-r_2+m_1-m_2+1}
\eea
With this form we have 
\bea
{\rm Tr}(P^{n,m,0}_{R,(r,s)}\, \hat{C})=
{d_r\chi_s ((m,m-1))(m-1)+n\chi_{R,(r,s)}(m,m+1)-(r_2+m_2-2)d_r d_s\over r_1-r_2+2j_3+1}\label{AAident}
\eea
This is very similar to what we had before, and we see that the same restricted character appears.
A new feature is the appearance of an $S_m$ character, which is easily evaluated using the Murnaghan-Nakayama
rule\cite{hammer} (recall that $s=2j$)
\bea
   \chi_s ((m,m-1))=d_s {2j+2j^2+{m\over 2}(m-4)\over m(m-1)}
\eea
This is everything we need to evaluate (\ref{AAident}).
In the end we find
\bea
   \sum_{s^-}{\rm Tr}(P^{n,m,0}_{R,(r,s)}\, P^{n,m-1,1}_{R,(r,s^-,{\tiny \yng(1)})})
   =d_r d_s\left[ {m+2j_3\over 2m}+{2j(j+1)-2j_3^2-m\over 2m(r_1-r_2+2j_3+1)}\right]\, .\label{scndsu3}
\eea
Remarkably, these are exactly the same identities we had from the $su(2)$ problem, except that $r_1-r_2$ is replaced
with $r_1-r_2+2j_3$, which is very natural.
There are only two non-zero terms on the LHS of the identities (\ref{frstsu3}) and (\ref{scndsu3}).

\subsection{Simple traces}

There are a number of traces that are simple enough that they can be computed immediately.
The traces
\bea
{\rm Tr}(P_{R,(r,j,j,k,k)}^{n,m,p}P_{R,(r,j-1,j-1,k+1,k+1)}^{n,m-1,p+1})=d_r
\eea
\bea
{\rm Tr}(P_{R,(r,j,-j,k,-k)}^{n,m,p}P_{R,(r,j-1,-j+1,k+1,-k-1)}^{n,m-1,p+1})=d_r
\eea
follow because the subspaces we project to are the same for both projectors appearing in the trace.
For the special case that $r_1=r_2$, the $Y$ boxes are already organized in an irrep so that we reduce to the formulas
we obtained for the $su(2)$ case.
The traces we obtain in this way are as follows
\bea
{\rm Tr}(P_{R,(r,j,j,k,k_3)}^{n,m,p}P_{R,(r,j-{1\over 2},j-{1\over 2},k-{1\over 2},k_3+{1\over 2})}^{n,m+1,p-1})
=d_r d_s d_{t^-} \left[{k-k_3\over 2k}+{k^2-k_3^2\over 2k (2j+1)}\right]\,,
\eea
\bea
{\rm Tr}(P_{R,(r,j,j,k,k_3)}^{n,m,p}P_{R,(r,j-{1\over 2},j-{1\over 2},k+{1\over 2},k_3+{1\over 2})}^{n,m+1,p-1})
=d_r d_s d_{t^-} \left[{k+k_3+1\over 2k+2}-{(k+1)^2-k_3^2\over (2k+2) (2j+1)}\right]\,,
\eea
\bea
{\rm Tr}(P_{R,(r,j,j,k,k_3)}^{n,m,p}P_{R,(r,j+{1\over 2},j+{1\over 2},k-{1\over 2},k_3-{1\over 2})}^{n,m+1,p-1})
=d_r d_s d_{t^-} \left[{k+k_3\over 2k}-{k^2-k_3^2\over 2k (2j+1)}\right]\,,
\eea
\bea
{\rm Tr}(P_{R,(r,j,j,k,k_3)}^{n,m,p}P_{R,(r,j+{1\over 2},j+{1\over 2},k+{1\over 2},k_3-{1\over 2})}^{n,m+1,p-1})
=d_r d_s d_{t^-} \left[{k-k_3+1\over 2k+2}+{(k+1)^2-k_3^2\over (2k+2) (2j+1)}\right]\,.
\eea
We now have enough to write the general results.

\subsection{Exact Result for the $su(3)$ generators}

The exact results for the traces we need are
\bea
{\rm Tr}(P^{n,m,p}_{r_1,j,j_3,k,k_3}P^{n,m+1,p-1}_{r_1,j+{1\over 2},j_3+{1\over 2},k-{1\over 2},k_3-{1\over 2}})
=d_r d_s d_{t^-}\left[{k+k_3\over 2k}-{k^2-k_3^2\over 2k(r_1-r_2+2j_3+1)}\right]\cr
\times\left[{j+j_3+1\over 2j+1}+
{(j+{1\over 2})^2-(j_3+{1\over 2})^2\over (2j+1)(r_1-r_2+2j_3+2)}\right]
\eea
\bea
{\rm Tr}(P^{n,m,p}_{r_1,j,j_3,k,k_3}P^{n,m+1,p-1}_{r_1,j-{1\over 2},j_3+{1\over 2},k-{1\over 2},k_3-{1\over 2}})
=d_r d_s d_{t^-}\left[{k+k_3\over 2k}-{k^2-k_3^2\over 2k(r_1-r_2+2j_3+1)}\right]\cr
\times\left[{j-j_3\over 2j+1}-
{(j+{1\over 2})^2-(j_3+{1\over 2})^2\over (2j+1)(r_1-r_2+2j_3+2)}\right]
\eea
\bea
{\rm Tr}(P^{n,m,p}_{r_1,j,j_3,k,k_3}P^{n,m+1,p-1}_{r_1,j+{1\over 2},j_3+{1\over 2},k+{1\over 2},k_3-{1\over 2}})
=d_r d_s d_{t^-}\left[{k-k_3+1\over 2k+2}+{(k+1)^2-k_3^2\over (2k+2)(r_1-r_2+2j_3+1)}\right]\cr
\times\left[{j+j_3+1\over 2j+1}+
{(j+{1\over 2})^2-(j_3+{1\over 2})^2\over (2j+1)(r_1-r_2+2j_3+2)}\right]\cr
\eea
\bea
{\rm Tr}(P^{n,m,p}_{r_1,j,j_3,k,k_3}P^{n,m+1,p-1}_{r_1,j-{1\over 2},j_3+{1\over 2},k+{1\over 2},k_3-{1\over 2}})
=d_r d_s d_{t^-}\left[{k-k_3+1\over 2k+2}+{(k+1)^2-k_3^2\over (2k+2)(r_1-r_2+2j_3+1)}\right]\cr
\times\left[{j-j_3\over 2j+1}-
{(j+{1\over 2})^2-(j_3+{1\over 2})^2\over (2j+1)(r_1-r_2+2j_3+2)}\right]\cr
\eea
\bea
{\rm Tr}(P^{n,m,p}_{r_1,j,j_3,k,k_3}P^{n,m+1,p-1}_{r_1,j+{1\over 2},j_3-{1\over 2},k-{1\over 2},k_3+{1\over 2}})
=d_r d_s d_{t^-}\left[{k-k_3\over 2k}+{k^2-k_3^2\over 2k(r_1-r_2+2j_3+1)}\right]\cr
\times\left[{j-j_3+1\over 2j+1}-
{(j+{1\over 2})^2-(j_3+{1\over 2})^2\over (2j+1)(r_1-r_2+2j_3)}\right]
\eea
\bea
{\rm Tr}(P^{n,m,p}_{r_1,j,j_3,k,k_3}P^{n,m+1,p-1}_{r_1,j-{1\over 2},j_3-{1\over 2},k-{1\over 2},k_3+{1\over 2}})
=d_r d_s d_{t^-}\left[{k-k_3\over 2k}+{k^2-k_3^2\over 2k(r_1-r_2+2j_3+1)}\right]\cr
\times\left[{j+j_3\over 2j+1}+
{(j+{1\over 2})^2-(j_3-{1\over 2})^2\over (2j+1)(r_1-r_2+2j_3)}\right]
\eea
\bea
{\rm Tr}(P^{n,m,p}_{r_1,j,j_3,k,k_3}P^{n,m+1,p-1}_{r_1,j+{1\over 2},j_3-{1\over 2},k+{1\over 2},k_3+{1\over 2}})
=d_r d_s d_{t^-}\left[{k+k_3+1\over 2k+2}-{(k+1)^2-k_3^2\over (2k+2)(r_1-r_2+2j_3+1)}\right]\cr
\times\left[{j-j_3+1\over 2j+1}-
{(j+{1\over 2})^2-(j_3-{1\over 2})^2\over (2j+1)(r_1-r_2+2j_3)}\right]\cr
\eea
and
\bea
{\rm Tr}(P^{n,m,p}_{r_1,j,j_3,k,k_3}P^{n,m+1,p-1}_{r_1,j-{1\over 2},j_3-{1\over 2},k+{1\over 2},k_3+{1\over 2}})
=d_r d_s d_{t^-}\left[{k+k_3+1\over 2k+2}-{(k+1)^2-k_3^2\over (2k+2)(r_1-r_2+2j_3+1)}\right]\cr
\times\left[{j+j_3\over 2j+1}+
{(j+{1\over 2})^2-(j_3-{1\over 2})^2\over (2j+1)(r_1-r_2+2j_3)}\right]\, .\cr
\eea
Using these results we find that the exact matrix elements of ${\rm Tr}\left( Y{d\over dX}\right)$ are
\bea
{\rm Tr}\left( Y{d\over dX}\right)O^{n,m,p}_{r_1,j,j_3,k,k_3}=
\sum_{a,b,c,d=-{1\over 2}}^{1\over 2}C(a,b,c,d)\,
O^{n,m+1,p-1}_{r_1,j+a,j_3+b,k+c,k_3+d}
\eea
where
\bea
C({1\over 2},{1\over 2},-{1\over 2},-{1\over 2})
=\sqrt{{p+2k+2\over 2}{2k\over 2k+1}}\left[{k+k_3\over 2k}-{k^2-k_3^2\over 2k(r_1-r_2+2j_3+1)}\right]\cr
\times
\sqrt{{m+2j+4\over 2}{2j+1\over 2j+2}}
\left[{j+j_3+1\over 2j+1}+
{(j+{1\over 2})^2-(j_3+{1\over 2})^2\over (2j+1)(r_1-r_2+2j_3+2)}\right]
\eea
\bea
C(-{1\over 2},{1\over 2},-{1\over 2},-{1\over 2})
=\sqrt{{p+2k+2\over 2}{2k\over 2k+1}}\left[{k+k_3\over 2k}-{k^2-k_3^2\over 2k(r_1-r_2+2j_3+1)}\right]\cr
\times\sqrt{{m-2j+2\over 2}{2j+1\over 2j}}
\left[{j-j_3\over 2j+1}-
{(j+{1\over 2})^2-(j_3+{1\over 2})^2\over (2j+1)(r_1-r_2+2j_3+2)}\right]
\eea
\bea
C({1\over 2},{1\over 2},{1\over 2},-{1\over 2})
=\sqrt{{p-2k\over 2}{2k+2\over 2k+1}}\left[{k-k_3+1\over 2k+2}+{(k+1)^2-k_3^2\over (2k+2)(r_1-r_2+2j_3+1)}\right]\cr
\times\sqrt{{m+2j+4\over 2}{2j+1\over 2j+2}}
\left[{j+j_3+1\over 2j+1}+
{(j+{1\over 2})^2-(j_3+{1\over 2})^2\over (2j+1)(r_1-r_2+2j_3+2)}\right]\cr
\eea
\bea
C(-{1\over 2},{1\over 2},{1\over 2},-{1\over 2})
=\sqrt{{p-2k\over 2}{2k+2\over 2k+1}}\left[{k-k_3+1\over 2k+2}+{(k+1)^2-k_3^2\over (2k+2)(r_1-r_2+2j_3+1)}\right]\cr
\times\sqrt{{m-2j+2\over 2}{2j+1\over 2j}}\left[{j-j_3\over 2j+1}-
{(j+{1\over 2})^2-(j_3+{1\over 2})^2\over (2j+1)(r_1-r_2+2j_3+2)}\right]\cr
\eea
\bea
C({1\over 2},-{1\over 2},-{1\over 2},{1\over 2})
=\sqrt{{p+2k+2\over 2}{2k\over 2k+1}}\left[{k-k_3\over 2k}+{k^2-k_3^2\over 2k(r_1-r_2+2j_3+1)}\right]\cr
\times\sqrt{{m+2j+4\over 2}{2j+1\over 2j+2}}
\left[{j-j_3+1\over 2j+1}-
{(j+{1\over 2})^2-(j_3+{1\over 2})^2\over (2j+1)(r_1-r_2+2j_3)}\right]
\eea
\bea
C(-{1\over 2},-{1\over 2},-{1\over 2},{1\over 2})
=\sqrt{{p+2k+2\over 2}{2k\over 2k+1}}\left[{k-k_3\over 2k}+{k^2-k_3^2\over 2k(r_1-r_2+2j_3+1)}\right]\cr
\times\sqrt{{m-2j+2\over 2}{2j+1\over 2j}}\left[{j+j_3\over 2j+1}+
{(j+{1\over 2})^2-(j_3-{1\over 2})^2\over (2j+1)(r_1-r_2+2j_3)}\right]
\eea
\bea
C({1\over 2},-{1\over 2},{1\over 2},{1\over 2})
=\sqrt{{p-2k\over 2}{2k+2\over 2k+1}}\left[{k+k_3+1\over 2k+2}-{(k+1)^2-k_3^2\over (2k+2)(r_1-r_2+2j_3+1)}\right]\cr
\times\sqrt{{m+2j+4\over 2}{2j+1\over 2j+2}}
\left[{j-j_3+1\over 2j+1}-
{(j+{1\over 2})^2-(j_3-{1\over 2})^2\over (2j+1)(r_1-r_2+2j_3)}\right]\cr
\eea
and
\bea
C(-{1\over 2},-{1\over 2},{1\over 2},{1\over 2})
=\sqrt{{p-2k\over 2}{2k+2\over 2k+1}}\left[{k+k_3+1\over 2k+2}-{(k+1)^2-k_3^2\over (2k+2)(r_1-r_2+2j_3+1)}\right]\cr
\times\sqrt{{m-2j+2\over 2}{2j+1\over 2j}}\left[{j+j_3\over 2j+1}+
{(j+{1\over 2})^2-(j_3-{1\over 2})^2\over (2j+1)(r_1-r_2+2j_3)}\right]\, .\cr
\eea
The complete set of generators can now easily be obtained by hermitian conjugation and by using the $su(3)$ algebra.
It is straight forwards to check the above traces numerically, to check that the generators reduce to the displaced corners
generators in the correct limit, to explicitly check their action for small values of $n,m,p$ and to verify that they close the
correct algebra. 
This achieves the aim of this article.

\section{Discussion}

In this article we developed techniques that allow the computation of the exact structure of infinitesimal $su(3)$
rotations of restricted Schur polynomials constructed using three complex adjoint scalar fields.
A key ingredient has been appreciating that there is a natural tensor product structure to the calculation which
reduces the problem to determining the decomposition of a projection operator under a subgroup.
This problem has been effectively handled using Jucys-Murphy elements and the same methods can be applied to determine
the action of $su(\cdot)$ global rotations acting on fermionic fields in restricted Schur operators constructed using both
fermions and bosons\cite{Koch:2012sf}.

Our main motivation in this article has been to explore the action of the dilatation operator, along the lines suggested 
in \cite{Koch:2013xaa}.
Indeed, using the displaced corners approximation \cite{Koch:2013xaa} constructed the $su(2)$ generators.
The requirement that these generators commute with the dilatation operator gives a set of recursion relations that can
be solved after some input from perturbative field theory is given.
The results reproduce the known one and two loop results and provide definite predictions for the higher loop structure
of the dilatation operator.
These results are in perfect harmony with expectations following from $su(2|2)$ symmetry \cite{Beisert:2005tm,Beisert:2006qh}
supporting their validity.  
The displaced corners approximation necessarily takes the number of $Y$ fields to be very much less than the number of
$Z$ field so that the analysis carried out in \cite{Koch:2013xaa} is necessarily restricted to small deformations of the half-BPS
sector.
The generators constructed in this paper are exact and thus the requirement that these generators commute with the dilatation
operator will lead to a new set of recursion relations that are generally valid.
A key insight of \cite{Koch:2013xaa} was that in the large $N$ limit, in the displaced corners approximation, $j_3$ is
conserved by the dilatation operator.
This conservation law was further identified with the absence of open string splitting at large $N$.
To solve the new set of recursion relations, it will be necessary to construct a similar understanding of how the simplicity of large $N$
manifests itself. 

The tensor product structure we have uncovered in this article is also present in operators constructed using the Brauer 
algebra \cite{Kimura:2007wy,Kimura:2009jf,Kimura:2012hp}.
The lessons learned in this article should be valuable for the construction of the action of global symmetry generators
on the Brauer basis too.
The construction of the global symmetry generators when acting on the basis developed in \cite{Brown:2007xh,Brown:2008ij}
is likely to be straight forward, given the fact that this basis is constructed so that it has good global quantum numbers.
The construction of these generators may provide an approach to the study of the spectrum of the dilatation operator
in these bases, an important problem that has not been explored much. 

{\vskip 1.0cm}

\noindent
{\it Acknowledgements:} 
This work is supported by the South African Research Chairs
Initiative of the Department of Science and Technology and the National Research Foundation.
Any opinion, findings and conclusions or recommendations expressed in this material
are those of the authors and therefore the NRF and DST do not accept any liability
with regard thereto.
Research at Perimeter Institute is supported by the Government of Canada
through Industry Canada and by the Province of Ontario through the Ministry of Economic Development \&
Innovation.

\end{document}